%% file: Canellas_UnsafeAtAnyLevel_Resubmission2.tex
\documentclass[acmtog, authorversion]{acmart}

\usepackage{booktabs} % For formal tables

\setcitestyle{square}

\usepackage{hyperref}
\usepackage[ruled]{algorithm2e} % For algorithms

\SetAlFnt{\small}
\SetAlCapFnt{\small}
\SetAlCapNameFnt{\small}
\SetAlCapHSkip{0pt}
\IncMargin{-\parindent}

\acmDOI{10.1145/3342102}

\begin{document}
% Title portion
\title{Unsafe At Any Level}
%\titlenote{This is a titlenote}
\subtitle{NHTSA's levels of automation are a liability for autonomous vehicle design and regulation}
%\subtitlenote{Subtitle note}

\author{Marc Canellas}
\orcid{0000-0002-2952-1016}
\affiliation{%
  \institution{Vice-Chair, IEEE-USA AI Policy Committee; JD Candidate, New York University School of Law}
  \streetaddress{}
  \city{New York}
  \state{NY}
  \country{USA}}
\email{marc.c.canellas@gmail.com.}

\author{Rachel Haga}
\affiliation{%
  \institution{Member, IEEE-USA AI Policy Committee; Data Scientist, Elicit Insights LLC}
  \city{New York}
\state{NY}
\postcode{}
\country{USA}}
\email{rachel.haga@gmail.com}

%\affiliation{%
% \institution{Rajiv Gandhi University}
%\email{aprna_patel@rguhs.ac.in}
%\affiliation{%
%  \institution{Tsinghua University}
%  \streetaddress{30 Shuangqing Rd}
%  \country{China}
%}
%\email{chan0345@tsinghua.edu.cn}
%\author{Ting Yan}
%\affiliation{%
%  \institution{Eaton Innovation Center}
%  \city{Prague}
%  \country{Czech Republic}}
%\email{yanting02@gmail.com}
%\author{Tian He}
%\affiliation{%
%  \institution{University of Virginia}
%  \department{School of Engineering}
%  \city{Charlottesville}
%  \state{VA}
%  \postcode{22903}
%  \country{USA}
%}
%\affiliation{%
%  \institution{University of Minnesota}
%\email{tinghe@uva.edu}
%\author{Chengdu Huang}
%\author{John A. Stankovic}
%\author{Tarek F. Abdelzaher}
%\affiliation{%
%  \institution{University of Virginia}
%  \department{School of Engineering}
%  \city{Charlottesville}
%  \state{VA}
%  \postcode{22903}
%  \country{USA}
%}
%
%\renewcommand\shortauthors{Zhou, G. et al}
%%
\begin{abstract}
\textbf{This is a post-peer-review, pre-copyedit version of an article published in Communications of the ACM, March 2020, Vol. 63 No. 3, Pages 31-34. The final authenticated version is available online at: \url{http://dx.doi.org/10.1145/3342102}. The contents of this Viewpoint are solely the responsibility of the authors.}
\end{abstract}

%
%%
%% The code below should be generated by the tool at
%% http://dl.acm.org/ccs.cfm
%\begin{CCSXML}
%<ccs2012>
%  <concept_id>10010520.10010553.10010562</concept_id>
%  <concept_desc>Computer systems organization~Embedded systems</concept_desc>
%  <concept_significance>500</concept_significance>
%  <concept_id>10010520.10010575.10010755</concept_id>
%  <concept_desc>Computer systems organization~Redundancy</concept_desc>
%  <concept_significance>300</concept_significance>
%  <concept_id>10010520.10010553.10010554</concept_id>
%  <concept_desc>Computer systems organization~Robotics</concept_desc>
%  <concept_significance>100</concept_significance>
%  <concept_id>10003033.10003083.10003095</concept_id>
%  <concept_desc>Networks~Network reliability</concept_desc>
%  <concept_significance>100</concept_significance>
%</ccs2012>
%\end{CCSXML}
%
%\ccsdesc[500]{Computer systems organization~Embedded systems}
%% End generated code
%%
%
%
%\keywords{Wireless sensor networks, media access control,
%multi-channel, radio interference, time synchronization}
%
%

\maketitle

\input{samplebody-journals2}

\end{document}

%% file: samplebody-journals2.tex
Walter Huang, a 38-year-old Apple Inc. engineer, died on March 23, 2018, after his Tesla Model X crashed into a highway barrier in Mountain View, California.\footnote{See \url{https://www.bloomberg.com/news/articles/2018-04-12/a-timeline-of-the-tesla-autopilot-crash-investigation}.} Tesla immediately disavowed responsibility for the accident. ``The fundamental premise of both moral and legal liability is a broken promise, and there was none here: [Mr. Huang] was well aware that the Autopilot was not perfect [and the] only way for this accident to have occurred is if Mr. Huang was not paying attention to the road, despite the car providing multiple warnings to do so.''\footnote{See \url{https://assets.bwbx.io/documents/users/iqjWHBFdfxIU/rr5U4aZLWK5A/v0}.}

This is the standard response from Tesla and Uber, the manufacturers of the automated vehicles involved in the six fatal accidents to date: the automated vehicle isn’t perfect, the driver knew it wasn’t perfect, and if only the driver had been paying attention and heeded the vehicle’s warnings, the accident would never have occurred.\footnote{After fatal accidents in China and Florida in 2016, Tesla responded that ``every time the Autopilot is engaged, the car reminds the driver to `Always keep your hands on the wheel. Be prepared to take over at any time' '' (\url{https://www.tesla.com/blog/tragic-loss}). After a fatal accident in Arizona in March, 2018, Uber responded by installing new driver monitoring systems for detecting ``inattentive behavior'' (\url{https://www.theverge.com/2018/7/24/17607898/uber-self-driving-car-public-roads-driver-monitoring}).  After the fourth fatal Tesla accident 	in Delray Beach, Florida, in 2019, Tesla responded that ``when used properly by an attentive driver who is prepared to take control at all times, drivers supported by Autopilot are safer than those operating without assistance.'' (\url{https://www.forbes.com/sites/alanohnsman/2019/05/16/investigators-say-tesla-model-3-driver-killed-in-fl orida-crash-used-autopilot/}).} However, as researchers focused on human-automation interaction in aviation and military operations, we cannot help but wonder if there really are no broken promises and no legal liabilities.

These automated vehicle accidents are predicted by the science of human-automation interaction and the major aviation accidents caused, in large part, by na\"{i}ve implementation of automation in the cockpit and airspace. Aviation has historically been plagued by designers ignoring defects until they have caused fatal accidents. We even have a term for this attitude: tombstone design. Acknowledging tragedies and the need to better understand their causes led aviation to become the canonical domain for understanding human-automation interaction in complex, safety-critical operations. Today, aviation is an incredibly safe mode of transportation, but we are constantly reminded of why we must respect the realities of human-automation interaction. A recent tragic example is Boeing 737 MAX 8's MCAS automation which contributed to two crashes and the deaths of 346 people before the human-automation interaction failure was publicly acknowledged.

Science like human-automation interaction has a critical role in determining legal liability, and courts appropriately rely on scientists and engineers to determine whether an accident, or harm, was foreseeable. Specifically, a designer could be found liable if, at the time of the accident, scientists knew there was a systematic relationship between the accident and the designer’s untaken precaution \cite{Grady_2002_ProximateCauseDecoded}.

The scientific evidence is undeniable. There is a systematic relationship between the design of automated vehicles and the types of accidents that are occurring now and will inevitably continue to occur in the future. These accidents were not unforeseeable and the drivers were not exclusively to blame. In fact, the vehicle designs and fatalities are both symptoms of a larger failed system: the five levels of automation (LOA) for automated vehicles.

The LOA framework is defined in the SAE International J3016 Standard (SAE J3016)\cite{SAEInternational_2018_J3016TaxonomyDefinitions} and adopted as the U.S. National Highway Transportation Safety Administration's (NHTSA) standard automated vehicle categories \cite{NHTSA_2018_AutomatedVehicles}. The LOA framework is premised on the idea that automation is collaborating at various levels of interaction as part of a team with a human operator. The typical LOA is a one-dimensional spectrum of interaction ranging from fully-manual to fully-automated, exemplified by NHTSA's Level 0 and Level 5. For their part, SAE states that their LOA ``provides a logical taxonomy for [classification]... in order to facilitate clear communications'' and caveats that their LOA ``is not a specification and imposes no requirements'' \cite{SAEInternational_2018_J3016TaxonomyDefinitions}.

The central fl aw of LOA is right there in its name. Levels of \textit{automation} focus on a singular, static definition of the automation’s capabilities, ignoring the deeper ideas of teamwork, collaboration, and interdependency necessary for mission success – in this case operating a vehicle. Just reading the names of NHTSA’s levels, you can see that the focus is solely on what the automation can do: 0 - No Driving Automation; 1 - Driver Assistance; 2 - Partial Driving Automation; 3 - Conditional Driving Automation; 4 - High Driving Automation; 5 - Full Driving Automation.

This automation-centric perspective is counter to the idea of teamwork and explains why, despite their former prevalence in the academic literature, LOA is now acknowledged to be limited, problematic, and, to some, worth discarding altogether \cite{Feigh_2014_RequirementsEffectiveFunction,Hoffman_2013_SevenDeadlyMyths}.. Even Tom Sheridan, who originated the idea of LOA in 1978 \cite{Sheridan_1978_HumanComputerControl}, explained recently that LOA was never intended to be ``a prescription for designing automation'' and that the NHTSA's categories for automated vehicles is a key example of ``LOA that are not appropriate to [their] given context,'' not only in design but also in taxonomy and communication \cite{Sheridan2018}.\footnote{For a thorough discussion of the problems of current single-dimensional LOA and how they can be modified to account for human capabilities and needs, see the special issue on Advancing Models of Human-Automation Interaction in the Journal of Cognitive Engineering and Decision Making (\url{http://journals.sagepub.com/toc/edma/12/1}).}

The scientific literature shows that today's automated vehicles and corresponding LOA are characterized by the same serious design and communication flaws that human-automation interaction engineers have been fighting for nearly 70 years: automating as much as possible without concern for the human operator’s capabilities or needs; relying on hidden, interdependent and coupled tasks for safety; and requiring the operator to immediately take over control in emergency situations without explicit support.

To make some of these reasons more salient, imagine that you are part of a two-person team required to complete an assignment. Imagine that only your teammate was given the instructions for what was needed to complete the assignment. Conversely, you were only told that at some point your teammate may be unable to complete the assignment and, without prior notice, you will need to immediately finish it. You were also told that if your team fails to complete the assignment it is entirely your fault.

Is this a recipe for good teamwork and success? Would you feel the need to constantly monitor your teammate? Would you feel like you have all the responsibility for the outcome but limited or no ability to affect it? At what point would it be easier to just do the work on your own? 

With this example in mind, consider the definition of NHTSA’s Level 2 Partial Driving Automation. This is currently the highest level of automation allowed without formal regulation in many U.S. states and the level for each of the five fatal Tesla accidents.

\begin{quotation}
	SAE J3016 Level 2 Partial Driving Automation: The driving automation system (while engaged) performs part of the dynamic driving task by executing both the lateral and the longitudinal vehicle motion control subtasks, and disengages immediately upon driver request;
	
	The human driver (at all times) performs the remainder of the [dynamic driving task] not performed by the driving automation system; supervises the driving automation system and intervenes as necessary to maintain safe operation of the vehicle; determines whether/when engagement and disengagement of the driving automation system is appropriate; immediately performs the entire [dynamic driving task] whenever required or desired.
\end{quotation}

Level 2 is the first point where the automation assumes full control of the foundational ``lateral and longitudinal vehicle motion control subtasks'' typically performed by human drivers such as lane centering, parking assist, and adaptive cruise control. The fi rst stated role of the human driver in Level 2 is to ``(at all times) [perform] the remainder of the [dynamic driving task] not performed by the driving automation system.'' These remaining tasks include supervising the automation and intervening as necessary based on object and event detection.

This is where LOA begins to show itself to be inappropriate for design, taxonomy, or communication of the safety-critical aspects of human-automation interaction in driving contexts as alluded to by Sheridan \cite{Sheridan2018}. These remaining tasks are the textbook definition of leftover allocation: automate as many tasks as technology will permit and assume the human will pick up whichever tasks are leftover \cite{Bainbridge_1983_Ironiesautomation}. Leftover allocation often results in incoherent sets of tasks and situations where humans are being required to monitor automation or the environment for conditions beyond which the automation can operate \cite{Wiener_1980_Flightdeckautomation} -- situations in which humans are ineffective \cite{Molloy_1996_Monitoringautomatedsystem}.

Level 2 is oversimplifying and obscuring the interdependence of the human driver and the automated driving system, assuming that the human driver’s leftover tasks are complete, coherent, and capable of being performed. By focusing on ``who does what,'' instead of emphasizing ``how to work together,'' the LOA is giving ``the illusion that we can successfully deploy automation by simply assigning functions to automation that were once performed by people... [Neglecting] the fact that such assignments do not simply substitute automation for people but create new functions for the people who are left to manage the automation'' \cite{Lee2018}.

Level 2's distribution of tasks is particularly troubling because engineers have known since the 1950s that monitoring is not a task humans can maintain for extended periods of time \cite{Fitts1951}. When a driver's interactions are limited to monitoring, they will lose real-time situation awareness, which can result in surprises. Workload will spike during off -nominal situations and be excessively low during normal operations between spikes, ultimately leading to humans who are notionally ``in-the-loop'' becoming, practically, ``out-of-the-loop'' \cite{Bainbridge_1983_Ironiesautomation}. These spikes and lulls in workload can lead to the well-recognized problem of automation bias where humans will tend to disregard or not search for contradictory information in light of an automated judgment or decision that is accepted as correct \cite{Parasuraman_1997_Humansandautomation:}. Beyond automation bias, the lack of system interaction over a prolonged period prevents the human from acquiring expertise in the first place and can lead to long-term knowledge and skill degradation \cite{Feigh_2014_RequirementsEffectiveFunction}. Combining this degradation with an incoherent set of leftover tasks will make it all but impossible for a driver to make an informed decision in an emergency situation.

The Level 2 Partial Automation Vehicle standard concludes with a final, fatal flaw: requiring the human operator to determine ``whether/when engagement and disengagement of the driving automation system is appropriate,'' and if disengagement is necessary, ``immediately [perform] the entire [dynamic driving task].'' In complex work environments such as automated vehicles where many tasks are interdependent and hidden, the driver is unlikely to know when disengagement is ``appropriate'' -- especially given the ambiguity built into the SAE standard.\footnote{Two notable stipulations in the SAE standard expand the number and uncertainty of vehicles states that the driver would be required to monitor. By definition, ``Levels are \textit{assigned, rather than measured,} and reflect the \textit{design intent} for the driving automation system feature as defined by its manufacturer'' (8.2, emphasis added). Even further, the standard states that a system can deliver multiple features at different levels under varying conditions (8.4).} Studies have shown that these hidden interdependencies can result in insufficient coordination and exacerbate workload lulls and spikes \cite{Feigh_2014_RequirementsEffectiveFunction}. This makes for a prototypically \textit{brittle} human-automated system because there is no discussion of how the human operator should be supported during disengagement or takeover in emergency situations \cite{Norman_1990_Theproblemwithautomation:inappropriate}.

With this extensive history of human-automation interaction science we can now perform the foreseeability analysis the law requires: Is there existing scientific evidence for a relationship between the accidents like the one that killed Mr. Huang and the design of Level 2 Partial Automation Vehicles?

In short, yes. Nearly 70 years of research argues against depending on human supervision of automation in complex, safety-critical environments without express consideration of the interdependent capabilities and needs of both the automation \textit{and} the human. It is insufficient, inappropriate, and dangerous to automate everything you can and leave the rest to the human. It is insufficient, inappropriate, and dangerous for NHTSA to allow automated vehicles to be designed this way.

Beyond the research, consider the paradoxical expectations for drivers who purchase and operate these automated vehicles. Drivers are sold the fantasy of being a passenger at times\footnote{A survey of 1,212 owners of automated vehicles revealed that the ``prevalence of drivers’ willingness to engage in other activities, look away from the roadway or rely on the technology to the exclusion of ordinary safe driving practices... may indicate lack of understanding or appreciation of the fact that these technologies are designed to assist the driver, and that the driver is still required to be attentive and in control of the vehicle at all times to ensure safety'' \cite{McDonald_2018_VehicleOwnersExperiences}.} but to the manufacturer they never stopped being the fully-liable driver.

NHTSA seems to have acknowledged the surface of these issues by providing human factors design guidance for Levels 2 and 3 because ``safe and efficient operation... requires [vehicles] be designed in a manner consistent with driver limitations, capabilities, and expectations'' \cite{Campbell2018}. However, this NHTSA guidance does not address the fundamental crisis of confidence in the LOA framework: can LOA appropriately regulate operations in complex work environments like automated vehicles \cite{Jamieson2018,Lee2018}? Does NHTSA's LOA simply need to be implemented better? Or does NHTSA need to completely reimagine their framework beyond LOA’s who-does-what perspective?

To answer this question, NHTSA should follow its own advice that ``lessons learned through the aviation industry’s experience with the introduction of automated systems may be instructive and inform the development of thoughtful, balanced approaches'' \cite{NHTSA_2018_AutomatedVehicles}. In 1989, in response to high-profile fatal accidents, the Air Transport Association of America (ATA) established a task force to examine the impact of automation on aviation safety. Their prescient conclusion remains true today  \cite{Billings_1997_Aviationautomation:search}:

\begin{quotation}
	During the 1970s and early 1980s... the concept of automating as much as possible was considered appropriate. The expected benefits were a reduction in pilot workload and increased safety... Although many of these benefits have been realized, serious questions have arisen and incidents/accidents have occurred which question the underlying assumption that maximum available automation is always appropriate or that we understand how to design automated systems so that they are fully compatible with the capabilities and limitations of the humans in the system.
\end{quotation}

Designers of automated vehicles face the same decisions today that aircraft designers have faced for decades. Automation has the potential to bring all the benefits of safety, reliability, economy, and comfort to our roads that have been brought to our airspace. But vehicle designers like Tesla and regulators like the NHTSA cannot abdicate their responsibility to stop foreseeable and preventable accidents by blaming the driver any more than aircraft designers can blame pilots. Aviation has already learned that tragedy should not be the only time regulations and designs are reconsidered. As automated vehicles begin driving in public spaces, entrusted with the lives of drivers, passengers, and pedestrians, these vehicle designers and regulators must learn from aviation’s tragic history of tombstone design, rather than repeating it.

~~~~

% Bibliography
\bibliographystyle{ACM-Reference-Format}
\bibliography{CACM}